\documentclass[11pt,twoside,natbib]{article}

\usepackage{asp2006}
\usepackage{epsf}
\usepackage{psfig}
\usepackage{lscape}
\usepackage{graphicx}
\def\lesssim{\mathrel{\hbox{\rlap{\hbox{ \lower4pt\hbox{$\sim$}}}\hbox{$<$}}}}

\def\gtrsim{\mathrel{\hbox{\rlap{\hbox{ \lower4pt\hbox{$\sim$}}}\hbox{$>$}}}}

\newcommand{\IRAS}{{\sl IRAS}}

\newcommand{\HST}{{\sl HST}}

\newcommand{\eg}{e.g.}

\newcommand{\bPic}{\hbox{$\beta$}~Pic}

\newcommand{\mdot}{{\mathrm M}_\odot}

\begin{document}
\title{Disks around Brown Dwarfs and Cool Stars}   
\author{D\'aniel Apai}   
\affil{Steward Observatory, The University of Arizona \\ 933 N. Cherry Avenue, Tucson, AZ 85721, and \\ Laplace Team, NASA Astrobiology Institute}    

\author{Kevin Luhman}   
\affil{Dept. of Astronomy \& Astrophysics, Pennsylvania State University\\
525 Davey Lab, University Park, PA 16802}    

\author{Michael C. Liu}   
\affil{Institute for Astronomy, University of Hawaii\\ 2680 Woodlawn Drive, Honolulu, HI 96822}    

\begin{abstract} 
We review the current picture of  disks around cool stars and brown dwarfs, including
disk fractions, mass estimates, disk structure and dispersal, accretion, dust composition, the debris disk phase. 
We discuss these in the framework of recent planet formation models. 
\end{abstract}


The study of disks around brown dwarfs and cool stars allows testing disk physics and planet formation processes  in a parameter range vastly different from those characteristic to the vicinity of Sun--like stars. 
This chapter provides a concise summary of seven reviews presented in the splinter session "Disks around Cool Stars and Brown Dwarfs" that placed the most recent results into an evolutionary picture.

\section{Disk Fractions and Disk Lifetimes}   

Disk fractions for young stars are measured through IR photometry of
young clusters, and these data are used to identify the stars that
have long wavelength excess emission that is indicative of disks.
Combining disk fractions for populations across a range of ages then
provides an estimate of the average disk lifetime.
Because the Spitzer Space Telescope is highly sensitive and can quickly
image young clusters, it can reliably and efficiently detect disks for brown
dwarfs at very low masses and for large numbers of brown dwarfs. For instance,
Luhman et al. (2005) used IRAC on Spitzer to obtain mid-IR images of
IC~348 and Chamaeleon~I, which encompassed 25 and 18 spectroscopically
confirmed low-mass members
of the clusters, respectively ($>$M6, $M<0.08$~$M_\odot$).
They found that $42\pm13$\% and $50\pm17$\% of the two samples exhibit
excess emission indicative of circumstellar disks.
In comparison, the disk fractions for stellar members of these clusters
are $33\pm4$\% and $45\pm7$\%
(M0-M6, 0.7~$M_\odot> M>0.1$~$M_\odot$).
The similarity of the disk fractions of stars and brown dwarfs
indicates that the raw materials for planet formation are available
around brown dwarfs as often as around stars and supports the notion
that stars and brown dwarfs share a common formation history.
However, as with the continuity of accretion rates from stars to brown dwarfs,
these results do not completely exclude some
scenarios in which brown dwarfs form through a distinct mechanism. For instance,
during formation through embryo ejection, the inner regions of disks
that emit at mid-IR wavelengths could survive, although one might expect
these truncated disks to have shorter lifetimes than those around stars.

When disk fractions for stellar populations across a range of ages
(0.5-30~Myr) are compared, they indicate that the inner disks around stars
have lifetimes of $\sim6$~Myr (Hillenbrand et al. 1998; Haisch et al. 2001).
The most accurate measurements of disk fractions for brown dwarfs are for
IC~348 and Chamaeleon~I, both of which have ages near 2~Myr, and so
a comparable estimate of the disk lifetime for brown dwarfs is not currently
possible. However, the presence of a disk around a brown dwarf
in the TW~Hya association
(Mohanty et al. 2003; Sterzik et al. 2004), which has an age
of 10~Myr, does suggest that the lifetime of brown dwarf disks might be at least
as long as that of stars. In fact, a preliminary measurement of the disk
fraction of low-mass members of Upper Sco (5~Myr) is similar to that found in 
IC~348 and Chamaeleon~I (Bouy et al. 2007), whereas the disk fraction of stars
is significantly lower in Upper Sco than in IC~348 (Lada et al. 2006;
Carpenter et al. 2006). Thus, disk lifetimes may be longer around
brown dwarfs than stars. Photometry and spectroscopy of a much larger
number of low-mass members of Upper Sco is currently ongoing with the
Spitzer Space Telescope, which should more definitively address this issue.

\section{Accretion in Substellar Disks}   

\begin{figure}[!ht]
\begin{center}
\includegraphics[width=6cm,height=6cm]{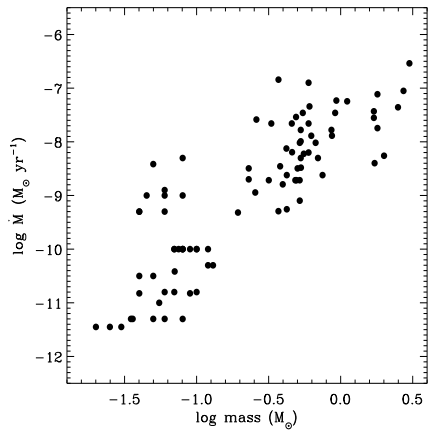}
\end{center}
\caption{{\small Mass accretion rate as a function of substellar and stellar mass for objects in Taurus (1~Myr), Cha~I (2~Myr), IC~348 (2~Myr), and Ophiuchus (0.5~Myr) (Gullbring et al. 1998; White \& Ghez 2001; Muzerolle et al. 2000, 2003, 2005; Natta et al. 2004; Mohanty et al. 2005). These regions exhibit similar accretion rates at a given mass, except for slightly higher rates in Ophiuchus.}}
\end{figure}

The first detection of accretion in an object near the hydrogen burning
limit was presented by Muzerolle et al. (2000) for V410~Anon~13 in Taurus
(Brice\~no et al. 2002).
The H$\alpha$ profile for this object exhibited an infall asymmetry
indicating ballistic infall at velocities consistent with the object's
mass and radius. Modeling of the profile yielded
$\dot{M} \sim 5 \times 10^{-12} \; \mdot \; yr^{-1}$,
much smaller than the average rate of $\sim10^{-8} \; \mdot \; yr^{-1}$
for solar-mass T Tauri stars (Gullbring et al. 1998).
Since the work on V410~Anon~13,
dozens of substellar accretors have been identified down to masses
approaching the deuterium burning limit and with ages from 1 to 10 Myr
(e.g., Jayawardhana et al. 2003; Mohanty et al. 2005;
Muzerolle et al. 2003, 2005), allowing systematic studies of the
accretion properties from stellar to substellar masses.
Estimates of $\dot{M}$ based on line profile modeling decrease with mass
down to $M\sim0.02$~$\mdot$ with a functional form of $\dot{M} \propto M^2$
(Muzerolle et al. 2003, 2005, Fig. 1).
The continuity of this correlation across the substellar limit
supports the idea that brown dwarfs form via fragmentation and collapse
of cloud cores in the same manner as stars.

\section{Disk Structure and Mineralogy}   

Flared disks (opening angle increasing with the disk radius) 
have been successful in explaining the infrared spectral energy distributions (SEDs) of disks around 
intermediate--mass stars and many sun--like stars (e.g. Kenyon \& Hartmann 1987). 
Mid--infrared studies of disks around brown dwarfs, however, may suggest a somewhat different picture. 
Disk models constrained by accurate ground--based mid--infrared and sub--millimeter observations 
(e.g.  Natta et al. 2002, Apai et al. 2002, Pascucci et al. 2003, Apai et al. 2004, Mohanty et al. 2004,  Sterzik et al. 2004, 
Apai et al. 2005, Scholz et al. 2006) found that flat or mildly flared disks often provide better fits to the data than flared disks.
This reduced flaring may be 
the result of the larger dust grains settling toward the disk mid--plane, a possible first step toward
planet formation (e.g. Apai et al. 2005, Scholz et al. 2006)


The 10~$\mu$m--window also offers a potential to probe the dust properties:
The strength and shape of the silicate emission bands is characteristic to the size, chemical composition
and lattice structure of the material. 
Broad--band photometry of the 10~$\mu$m silicate feature 
showed weak or absent silicate emission feature around six brown dwarfs, 
including 1--4 Myr--old Cha I members, $\rho$~Oph and Taurus members 
(Apai et al. 2002, 2004; Mohanty et al. 2004) and the 8~Myr--old TW Hy group member 2M1207 (Sterzik et al. 2004). 
These observations showed that the optically thin upper layers of these disks are often lacking sub--micron sized amorphous silicate grains, a major component of the interstellar medium. The absence of these grains suggested dust processing through collisional evolution.

Recently, Spitzer spectroscopy has been providing a new, detailed picture on the silicate emission feature for
an increasing number of disks around cool stars. Crystalline emission features has been identified 
by several groups (e.g. Forrest et al. 2004; Furlan et al. 2005; Apai et al. 2005;  Kessler-Silacci et al. 2006).
In a large sample of disks Kessler-Silacci et al. (2006) found an anti--correlation between the feature strength and  the spectral type of the stars, suggesting larger silicate grains around lower--mass stars. In a comparative, quantitative dust  composition study of Herbig Ae/Be, T Tauri and brown dwarf disks Apai et al. (2005) found that the crystalline mass fraction of the micron--sized  dust particles is higher around cooler central stars. Both these studies suggest that the observed dust is more processed around lower--mass stars than around higher--mass stars. However, as pointed out in Apai et al. (2005) and Kessler-Silacci et al. (2006) the mid--infrared observations probe different radii around stars with different luminosity. 
Nevertheless, larger disk samples and an extended wavelength range will allow in the near--future quantitative comparisons of the dust processing and help identify the role of radial mixing, stellar luminosity, and disk structure
in dust evolution.

\section{Disk Masses}   

The mass of circumstellar disks is one of the key parameters determining their potential to form 
planetary systems. Disk mass measurements require detecting the dust emission at optically thin
wavelengths, a major observational challenge for the tenuous disks around cool stars and brown dwarfs.
A second obstacle is the set of untested assumptions necessary to convert the dust emission to 
total disk mass. Nevertheless, the first surveys of disks around brown dwarfs demonstrated the 
observational feasibility of these studies providing the first insights on how disk mass scales 
with stellar mass.

Using a combined sub--millimeter and millimeter survey Klein et al. (2003) carried out the first
systematic search for optically thin dust emission from brown dwarf disks. The age of the targets spread
from 1--4 Myr--old disks in the Taurus star--forming region through the Pleiades ($\sim$100 Myr) to 
older brown dwarfs in the Solar neighborhood that may harbor debris disks. 
This survey has identified sub--millimeter emission from two young disks, allowing disk mass 
estimates and powerful constraints for the disk models. 

Recently, Scholz et al. (2006) carried out a sub--millimeter survey in the Taurus star--forming region. 
This large and sensitive study covered  20 young brown dwarfs and identified five new possible
detections of millimeter emission from circumstellar dust. Using standard assumptions, 
Scholz et al. (2006) derived probable disk masses ranging from 0.3 to 6.8 M$_{\rm Jup}$. 
The authors argue that these disk masses are difficult to explain with truncated disks, predicted to surround
brown dwarfs ejected early from their accreting envelope.

Our current understanding of disk masses is severely limited by four factors: 1) the very small 
number of firm millimeter disk detections; 2) the uncertain dust opacities; 3) the grain size
distribution that may vary from disk to disk;  and, 4) the gas--to--dust mass ratio, canonically assumed to be 100. 
In addition, the interpretation of the data  is further complicated from the strong bias introduced 
by the flux limits imposed by the sensitivity of the current detector technology. In spite of these uncertainties, the current data set --- assuming that the typical dust properties are independent of the stellar mass --- is consistent with a disk mass/star mass ratio that is similar (few percent) over a large range of stellar masses, from intermediate--mass stars through young Sun--like stars to the substellar regime.
The new generation of sub--millimeter receiver arrays will bring a marked improvement in the sample 
size and the sensitivity of the observations. However, the other three factors represent a serious limitation to our understanding of disk masses and may account for non--systematic uncertainties of two orders of magnitude.

\section{Transition Disks around Cool Stars}   

Disk dispersal sets the time available for planet formation and thus understanding this process 
is essential for understanding disk evolution and planet formation.
Disks around very low--mass stars offer a very different parameter set to study these 
processes. While the disk fraction studies discussed previously measure
the median disk lifetime and its dispersion, the study of a few individual objects in the phase
of dispersing their disks --- transition disks --- provide import insights on the mechanisms in work.
Transition disks have been identified around a dozen young Sun--like stars and 
low--mass stars (e.g. Calvet et al. 2002). These disks display a partly or fully 
evacuated inner cavity, outside of which they harbor an optically thick dust disk. Several
mechanisms have been invoked to account for the observed inner disk clearing, 
including UV--photoevaporation, dynamical clearing via gravitational interactions with
a forming giant planet, or grain growth beyond tens of micron size in the inner disk.

The recent discovery of a transition disk around a brown dwarf in the star--forming region
IC~348  (Muzerolle et al. 2006) demonstrated that inner disk clearing is a common 
phenomenon around a very broad range of stars. Using SED models Muzerolle et al. (2006)
estimated the inner disk hole size to be 0.5--1~AU, much larger than what could be explained by
magnetospheric truncation or dust sublimation alone. 
The photoevaporation models critically depend on the UV flux, which may originate from
the chromosphere or the accretion shocks.  As Muzerolle et al. points out the UV flux  
expected around brown dwarfs 
is most likely too low to account for the dispersal of the inner disk within the observed 
few Myr. 
Most recently, Spitzer IRS spectrum of this brown dwarf disk showed
some silicate emission feature, which would argue against dust evolution 
as the cause of the disk dispersal.
Another, possibly viable option is the gravitational influence of a forming super--Earth or giant planet; 
such a planet would have needed to form on time scales less than a few Myr, the age
of the IC 348 cluster. 

Sensitive Spitzer Space Telescope surveys are further increasing the number
of known transition disks around brown dwarfs. By pinpointing the mechanisms 
responsible for clearing out individual disks for a large sample and comparing
them to transition disks around more massive stars Spitzer is expected to 
provide a deeper understanding of the disk dispersal process.

\section{Planet Formation around Cool Stars}

The recent exoplanet discoveries suggest that although exoplanetary systems are frequent
around M--dwarfs, their architecture differs from those around F, G, and K--type stars.
Overall, preliminary statistics suggests that gas giant planets in the inner planetary systems 
are $\sim3\times$ less common than around Sun--like stars (e.g., Butler et al. 2006). The discovery of 
two Neptune--mass exoplanets (GJ 436; GJ581), three
Jupiter--mass planets (GJ 849; two in the GJ 876 system, e.g. Butler et al. 2006) and recently 
two  probable "super--Earths" (e.g., Beaulieu et al. 2006) poses several theoretical  challenges.

Laughlin et al. (2004) pointed out that the lower disk mass and longer orbital periods around
low--mass stars prolong the core accretion time beyond the typical disk 
lifetimes. This effect suppresses the efficiency of gas giant formation around M--dwarfs, also
confirmed by the more realistic simulations by Ida \& Lin (2005), which also
include the decreased accretion of gap opening--planets and subsequent orbital
migration. They find that while Jupiter--mass planets will be less frequent around M--dwarfs,
a second peak corresponding to isolated, close--in Neptune--mass planets will be more 
pronounced than around Sun--like stars. Kennedy, Kenyon \& Bromley (2006) points out
that the snowline -- that separates rocky and icy planetesimals -- moves inward throughout
the pre--main sequence evolution of the low--mass star, which effect will even further enhance
the frequency of close--in super--Earths.

An alternative, competing mechanism to explain planet formation is gravitational instability
in a marginally unstable disk. Boss (2006a) showed that such disks -- even when orbiting 
M--dwarf stars -- undergo rapid spiral arm formation, followed by the formation of massive clumps. 
Although the current simulations are unable to trace the clumps for time scales comparable to 
the lifetime of the disk, several arguments suggest
that the clumps became self--gravitating and will undergo collapse. These simulations demonstrate
that gravitationally unstable disks may be able to form gas giant planets with 4--7 AU semi--major 
axes, in contrast to the core accretion scenario.
Boss (2006b) suggests that such unstable disks, when formed in high--mass, dense stellar clusters,  are likely to be exposed to EUV/FUV irradiation that may quickly erode the contracting protoplanetary 
atmospheres. This mechanism may explain the existence of Uranus-- and Neptune--like 
exoplanets at larger radii.

The predictions of the core accretion and gravitational instability models differ for low--mass
stars and likely even more for brown dwarfs. The expanding period coverage of the ongoing
RV--searches and the increasing number of planets detected via microlensing and 
transit surveys will directly test the predictions of the different planet formation 
models in the very near future.

\section{Debris Disks around Cool Stars}

After dissipation of their primordial planet-forming disks of gas and
dust, many stars possess debris disks
(Backman \& Paresce 1993).  The dust in debris disks is
continually generated from collisions of larger bodies (comets and
asteroids) that are otherwise undetectable.  Debris disks represent the
extrasolar analogs of the asteroid belt and Kuiper Belt in our own solar
system.

Ground- and space-based photometric studies have identified over a
hundred debris disks around AFGK-type dwarfs from their thermal dust
emission (e.g., Zuckerman 2001),
Greaves \& Wyatt 2003, Beichman et al. 2005,
Rhee et al. 2006).  However, little is known about debris
disks around M~dwarfs, as only a handful of examples have been found
from integrated-light searches (e.g., Song et al. 2002,
Liu et al. 2004, Low et al. 2005,
Lestrade {et~al.} 2006).  Past debris-disk searches have mostly
neglected and/or overlooked low-mass stars, largely due to sensitivity
limitations.  M~dwarfs have $>$10--1000$\times$ lower luminosities
compared to other (AFGK-type) debris disk host stars, and thus the
thermal continuum emission from circumstellar dust is expected to be
significantly fainter.  (In a similar fashion, one would expect that the
thermal emission of debris disks around brown dwarfs will be even more
difficult to detect, if such objects exist.)  For instance, \IRAS\ could
only detect the photospheres of a handful of nearby M~dwarfs at
60~\micron.  In addition, stellar winds from M~dwarfs might rapidly
remove circumstellar dust, leading to much smaller IR excesses compared
to higher mass stars (Plavchan {et~al.} 2005).

The potential value of these systems is demonstrated by the young
($\approx$12~Myr) star AU~Mic, the first robustly identified M~dwarf
debris disk system (Song et~al. 2002, Liu et al. 2004).  At
a distance of only 10~pc, its disk is seen in scattered light as far as
20\arcsec\ in radius (Kalas, Liu \& Matthews 2004).  Adaptive optics and
\HST\ imaging of the AU~Mic disk achieves a spatial resolution as good
as 0.4~AU (Liu 2004, Krist et al. 2005,  Metchev et al. 2005, Graham et al. 2007).  A rich variety of
substructure is observed, including multiple clumps and a large-scale
warp, and is suggestive of planetary companions.  The overall disk
surface brightness profile follows a broken power-law, with a relatively
shallow slope inside of $\approx$40~AU and a much steeper slope at
larger distances; this can be attributed to the combined dynamical
effects of drag forces (from the Poynting-Robertson effect and stellar
winds) and radiation pressure blowout of small grains
({Augereau} \& {Beust} 2006, {Strubbe} \& {Chiang} 2006).  

Thanks to its proximity to Earth and its status as the first large
scattered-light disk (in angular extent) found since the discovery of
the \bPic\ disk more than 20 years ago, the AU~Mic disk has been the
subject of intense recent study, with more than a dozen refereed papers
since its discovery.  Indeed, the AU~Mic disk bears a striking degree of
similarity to the \bPic\ disk in its properties and features; given that
these two systems are both members of the \bPic\ moving group, the
common age and birth environment of the two stars provides a means to
study the degree of synchronicity (or divergence) in disk evolution
(e.g. Liu 2004).  However, thus far AU Mic is the
singular example of a resolved debris disk around a low-mass star.
Ongoing searches with space-based and ground-based facilities (\eg,
Spitzer, AKARI, JCMT/SCUBA-2, and Herschel) offer the opportunity to
find more such systems amenable to scruntiny.



\acknowledgements 
{\small We are grateful to the speakers of the splinter session for their excellent reviews on planet formation (Alan Boss), 
substellar accretion (Subhanjoy Mohanty), transition disks (James Muzerolle) and disk masses (Alexander Scholz). We thank the participants of the Cool Stars 14 for taking such an active role in the discussions and the local and scientific  organizing committees for making this both a productive and enjoyable meeting.}




\end{document}